\documentclass[iop,useAMS,usenatbib]{emulateapj}

\usepackage{graphicx}
\usepackage{multirow}
\usepackage{color}
\usepackage{rotating}

\newcommand{\msun}{${\rm M_{\sun}}$}

\def\ltsima{$\; \buildrel < \over \sim \;$}
\def\simlt{\lower.5ex\hbox{\ltsima}}
\def\gtsima{$\; \buildrel > \over \sim \;$}
\def\simgt{\lower.5ex\hbox{\gtsima}}
\def\kms{{\rm\,km\,s^{-1}}}

\def\kpc{{\rm\,kpc}}

\def\msun{{\rm\,M_\odot}}

\def\pc{{\rm\,pc}}

\def\s{\ifmmode \widetilde \else \~\fi}
\def\={\overline}

\def\spose#1{\hbox to 0pt{#1\hss}}

\def\lta{\mathrel{\spose{\lower 3pt\hbox{$\mathchar"218$}}
     \raise 2.0pt\hbox{$\mathchar"13C$}}}
\def\gta{\mathrel{\spose{\lower 3pt\hbox{$\mathchar"218$}}
     \raise 2.0pt\hbox{$\mathchar"13E$}}}
\def\Dt{\spose{\raise 1.5ex\hbox{\hskip3pt$\mathchar"201$}}}    % upper case
\def\dt{\spose{\raise 1.0ex\hbox{\hskip2pt$\mathchar"201$}}}    % lower case

\def\dotsfill{\leaders\hbox to 1em{\hss.\hss}\hfill}

\def\ltsima{$\; \buildrel < \over \sim \;$}
\def\gtsima{$\; \buildrel > \over \sim \;$}
\def\lsim{\lower.5ex\hbox{\ltsima}}
\def\gsim{\lower.5ex\hbox{\gtsima}}
\def\lapp{\ifmmode\stackrel{<}{_{\sim}}\else$\stackrel{<}{_{\sim}}$\fi}
\def\gapp{\ifmmode\stackrel{>}{_{\sim}}\else$\stackrel{<}{_{\sim}}$\fi}

\shorttitle{Newtonian vs. MOND models of NGC 2419}
\shortauthors{Ibata et al.}

\begin{document}

\title{Polytropic model fits to the globular cluster NGC~2419 in Modified Newtonian Dynamics}

\author{R. Ibata\altaffilmark{1}, A. Sollima\altaffilmark{2}, C. Nipoti\altaffilmark{3}, M. Bellazzini\altaffilmark{4}, S.C. Chapman\altaffilmark{5}, E. Dalessandro\altaffilmark{3}}

\altaffiltext{1}{Observatoire Astronomique, Universit\'e de Strasbourg, CNRS, 11, rue de l'Universit\'e, F-67000 Strasbourg, France; rodrigo.ibata@astro.unistra.fr}

\altaffiltext{2}{INAF - Osservatorio Astronomico di Padova, vicolo
dell'Osservatorio 5, 35122, Padova, Italy}

\altaffiltext{3}{Dipartimento di Astronomia, Universit\`a degli Studi di Bologna, via Ranzani 1, I-40127 Bologna, Italy}

\altaffiltext{4}{INAF - Osservatorio Astronomico di Bologna, via Ranzani 1, 40127, Bologna, Italy}

\altaffiltext{5}{Institute of Astronomy, Madingley Road, Cambridge CB3 0HA, UK}

\begin{abstract}
We present an analysis of the globular cluster NGC~2419, using a polytropic model in Modified Newtonian Dynamics (MOND) to reproduce recently published high quality data of the structure and kinematics of the system. We show that a specific MOND polytropic model of NGC~2419 suggested by a previous study can be completely ruled out by the data. Furthermore, the highest likelihood fit polytrope in MOND is a substantially worse model (by a factor of $\sim 5000$) than a Newtonian Michie model studied in an earlier contribution by Ibata et al. (2011). We conclude that the structure and dynamics of NGC~2419 favor Newtonian dynamics and do indeed challenge the MOND theory.
\end{abstract}


\section{Introduction}
\label{sec:Introduction}

The well-known discrepancy between the luminous components of galaxies and their dynamics has been interpreted as evidence that vast quantities of some unknown type of dark (non-luminous) matter surround these celestial structures. Although this dark matter has not yet been directly detected, its existence is consistent with other astrophysical analyses, including the dynamics of galaxy clusters and the formation of large-scale structure \citep[see, e.g.,][]{Springel:2006p3642}. However, a very interesting alternative to this possibility, proposed by \citet{Milgrom:1983p15031}, is that our theory of gravity is incorrect or incomplete. According to this idea, General Relativity (or the Newtonian approximation to that theory) in the low-acceleration regime below a characteristic value $a_0 (\sim 1.2 \times 10^{-8}\, {\rm cm \, s^{-2}})$ under-predicts the actual acceleration. 

This Modified Newtonian Dynamics (MOND) theory has succeeded in passing numerous observational tests over the almost three decades since it was first proposed  (for instance, it is able to fit the rotation curves of low surface brightness galaxies, \citealt{McGaugh:1998p18778}, and  tidal dwarf galaxies, \citealt{Gentile:2007p18789}, and can fit gravitational lenses, \citealt{Shan:2008p18817}). Although there are outstanding problems with the predictions of the MOND theory (the growth of cosmological perturbations, \citealt{Dodelson:2006p17084}, the offset between lensing mass and baryons in the Bullet Cluster, \citealt{Clowe:2006p15213}, Solar System tests, \citealt{Milgrom:1983p15031,Sereno:2006p15936}, dynamical friction in dwarf galaxies, \citealt{Ciotti:2004p17054, SanchezSalcedo:2006p17060, Nipoti:2008p17050, Angus:2009p17066}, and the kinematics and density profile of satellite galaxies, \citealt{Klypin:2009p17870}, to list a few), the more widely-accepted Cold Dark Matter (CDM) paradigm has deep-set issues of its own (the missing satellite galaxies, the possible non-existence of dark matter cusps and the high angular momentum of galactic disks, \citealt{Binney:2004p17034, Primack:2009p17036})\footnote{Plausible simulated solutions for these problems with CDM have only recently been presented, yet they remain untested observationally.}. Determining which of the two competing theories of gravity represents reality remains a very important and fundamental task.

Recent improvements in the precision and multiplexing capabilities of spectrographs have led several teams to study large samples of stars in Galactic globular clusters as a means to test MOND on the scale of parsecs up to $\sim 100\pc$ \citep{Scarpa:2003p17001, Haghi:2009p16438, Jordi:2009p16532, Gentile:2010p16618, Lane:2010p16975}.
In a recent contribution to this effort, our team examined the distant halo globular cluster NGC~2419, re-measuring the surface brightness profile derived from HST/ACS, Subaru/Suprimecam and CFHT/MegaCam images, and obtaining high accuracy radial velocity measurements with the Keck/DEIMOS spectrograph \citep[][hereafter Paper~I]{Ibata:2011p18386}. We argued that this globular cluster is by far the best globular cluster target to test MOND with current instrumentation. Most importantly, it lies far enough from the Milky Way so as to relatively isolated, and so suffers little external acceleration. Furthermore, it is massive, so it offers a reasonable number of targets within reach of available high-resolution spectrographs, its velocity dispersion is large enough to be easily measurable, and it is spatially extended, meaning that a substantial fraction of the cluster lies in the low-acceleration regime where MOND differs from Newtonian gravity.

Outside of the very central regions, the cluster is circularly-symmetric in projection, furthermore, the cluster does not show evidence of significant rotation (Paper~I). These facts justify using spherical models to fit the system. The analysis presented in Paper~I  first examined Michie models, which are an extension to King models that allow for anisotropy in the orbits of the constituent stars of a spherical cluster. We found that a Michie model in Newtonian gravity (with a core radius of $r_c=11.7\pc$, a tidal radius of $r_t=331\pc$, a central velocity dispersion parameter of $\sigma_0=6.36\kms$, and an anisotropy radius of $r_a=1.5 \times r_c$) gives an excellent representation of the data, whereas Michie models in MOND have significantly lower likelihood (the best model being a factor of more than 40000 worse than the best model in Newtonian gravity).

We then expanded our analysis to examine a wider range of models, using a Markov-Chain Monte Carlo (MCMC) scheme to generate kinematic models consistent with the Jeans equation. The best such model in MOND was again found to be strongly disfavored compared to the best Michie model in Newtonian gravity.

However, these conclusions from Paper I were recently strongly criticized by \citet[][hereafter S11]{Sanders:2011p18268}, on the grounds that other models could reproduce the system well in MOND. The MCMC procedure mentioned above that we implemented in Paper~I examined one million different kinematic models. Nevertheless, it is certainly possible that a good model in MOND was missed by the MCMC routine. Indeed, S11 claims to have found a model that provides a counter-example refuting our conclusions, proposing a model in MOND that supposedly fits the available data. Here we will examine the particular polytropic model proposed by S11 and extend the analysis to search for the MOND polytropic model that best fits the data on NCG~2419.

\section{The polytropes}

\begin{figure}
\begin{center}
\includegraphics[angle=0, bb= 60 80 560 750, clip, width=8.5cm]{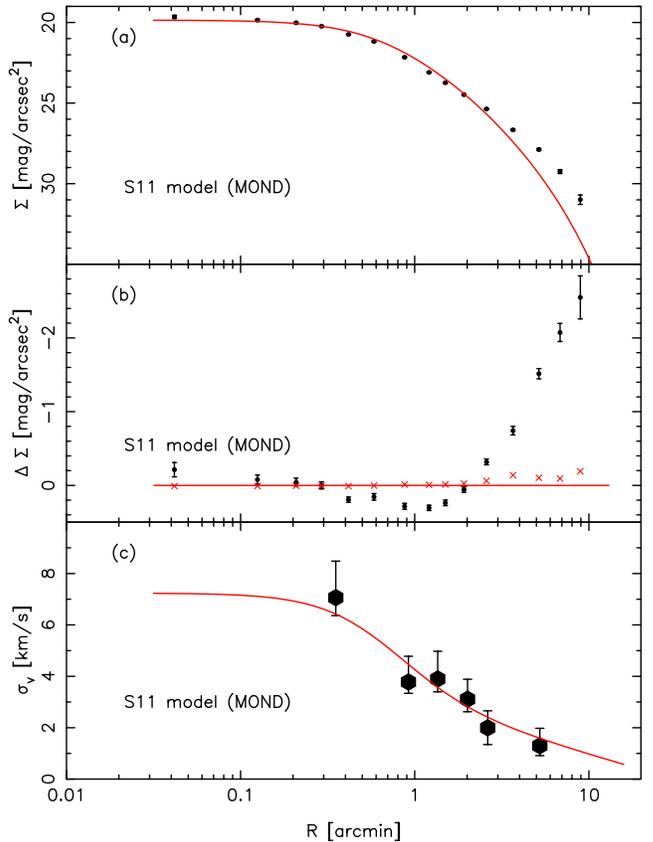}
\end{center}
\caption{Comparison of the surface brightness (top panel) and line of sight velocity dispersion (bottom panel) data from Paper~I to the S11 polytropic model (in MOND). The error bars mark 1-$\sigma$ uncertainties. The surface brightness measurements are derived from starcounts in annular regions containing between 104 and 910 stars, whereas the velocity dispersions are calculated from 27 stars per bin (except for the outermost bin which has 31 stars). The integrated mass of the model is $7.7\times10^5\msun$, and the mass to light ratio has been adjusted (to $M/L = 1.8$) to obtain the best $\chi^2$ fit to the surface brightness data. Nevertheless, the fit in panel (a) is exceedingly poor, having $\chi^2=1123$; this can be seen more clearly in the middle panel, where we have subtracted the model from panel (a). The cross markers show the values of model integrated over the same radial annuli as the data, taking into account the window function of the imaging survey. Clearly, this particular model is completely unacceptable.}
\label{fig:S11}
\end{figure}

The model proposed by S11 is a polytrope with polytropic index $n=10$, central velocity dispersion of $c_0=7.5 \kms$, and central density of $\rho_0=35 \msun \pc^3$. These parameters are related via the polytropic equation:
\begin{equation}
\overline{v_r^2} = c_0^2 (\rho/\rho_0)^{(1/n)} \, ,
\end{equation}
where $\overline{v_r^2}$ is the square of the radial velocity dispersion, and $\rho$ the density. In Paper~I, we found that a significant orbital anisotropy was necessary to reproduce the cluster satisfactorily. The model proposed by S11 also included anisotropy, modeled via the Osipkov-Merritt relation \citep{Osipkov:1979p15077,Merritt:1985p15033}
\begin{equation}
\beta(r)={{1}\over{1+(r_a/r)^2}} \, ,
\end{equation}
where $\beta \equiv 1 - \overline{v_\theta^2}/\overline{v_r^2}$ is the anisotropy parameter. The variable $\overline{v_\theta^2}$ is of course the square of the tangential (one-dimensional) velocity dispersion. As $r_a \rightarrow \infty$, the models become isotropic. The value adopted by S11 for the anisotropy radius was $r_a=18\pc$. Assuming that the cluster is spherical and static, the kinematics and structure must obey the spherical Jeans equation:
\begin{equation}
g = - {{\overline{v_r^2}}\over{r}} \Bigg[ {{\mathrm{d} \ln \rho}\over{\mathrm{d} \ln
        r}} + {{\mathrm{d} \ln \overline{v_r^2}}\over{\mathrm{d} \ln r}} + 2\beta \Bigg] \, ,
\label{eqn:Jeans}
\end{equation}
where $r$ is radial distance, and $g$ is the gravitational acceleration. It is straightforward to integrate this system of equations numerically, using a simple Euler scheme starting from $r=0$. This procedure works both for Newtonian dynamics and MOND, though to simulate the latter the acceleration $g$ needs to be altered according to the MOND prescription:
\begin{equation}
g \mu(g/a_0)=g_N
\end{equation}
where $g_N$ is the corresponding Newtonian acceleration that would result from the mass distribution given by $\rho(r)$, while $\mu$ is the MOND interpolation function. As S11, we take $\mu(x)=x/\sqrt{1+x^2}$, with $a_0=10^{-8} {\rm cm \, s^{-2}}$.

Figure~\ref{fig:S11} shows the surface brightness profile and line of sight velocity dispersion profile resulting from the parameters chosen by S11. The discrepancy with the observations is extremely large. From the 15 points in the surface brightness profile alone we obtain a chi-squared statistic of $\chi^2=1123$. While the S11 model is thus quantitatively completely excluded, it is very interesting nevertheless to examine whether other parameter values could provide an acceptable fit. To this end we used a general-purpose Markov-Chain Monte Carlo fitting algorithm to search the space of solutions of the polytropic models described above. The input parameters are $n$, $c_0$, $\rho_0$ and $r_a$, and the code sweeps through the solutions trying to find the most likely model. 

\begin{figure}
\begin{center}
\includegraphics[angle=0, bb= 60 80 560 750, clip, width=8.5cm]{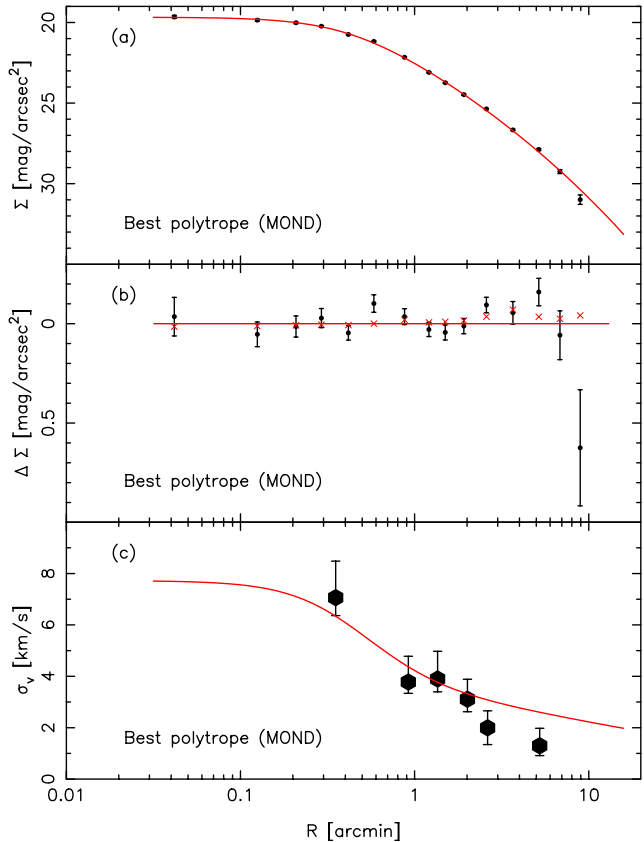}
\end{center}
\caption{As Figure~\ref{fig:S11}, but for the most likely stable polytropic model fit using the Markov-Chain Monte Carlo algorithm outlined in the text. MOND gravity is assumed. As before, the mass to light ratio has been adjusted to obtain the best $\chi^2$ fit to the surface brightness data. We require $M/L=1.9$ given the total model mass of $7.7\times 10^5 \msun$. However, this model provides a substantially worse fit to the data than the best Michie model reported in Paper~I (see Figure~\ref{fig:Michie}): the surface brightness model is a factor of 81 less likely, and the line of sight dispersion profile is a factor of 62 less likely. Note that the velocity dispersion data displayed in the bottom panel are shown only to guide the eye, as the likelihoods are calculated directly from the individual line of sight velocity measurements, with their corresponding (individual) uncertainty estimates.}
\label{fig:best}
\end{figure}

The (log) likelihood is calculated in a similar way to Paper~I, by adding the logarithm of the likelihood of the models given the surface brightness measurements to the logarithm of the likelihood of the models given the line of sight velocity measurements of individual stars. The surface brightness data are derived from starcounts measured in annuli over the surveyed area. As detailed in Paper~I, the HST and Subaru imaging of the cluster has significant gaps, so for any accurate model comparison it is necessary to take into account the missing areas. The window function of the imaging survey was therefore also applied to the models, and the models were integrated over exactly the same annuli as the data.

\begin{figure}
\begin{center}
\includegraphics[angle=0, bb= 60 80 560 750, clip, width=8.5cm]{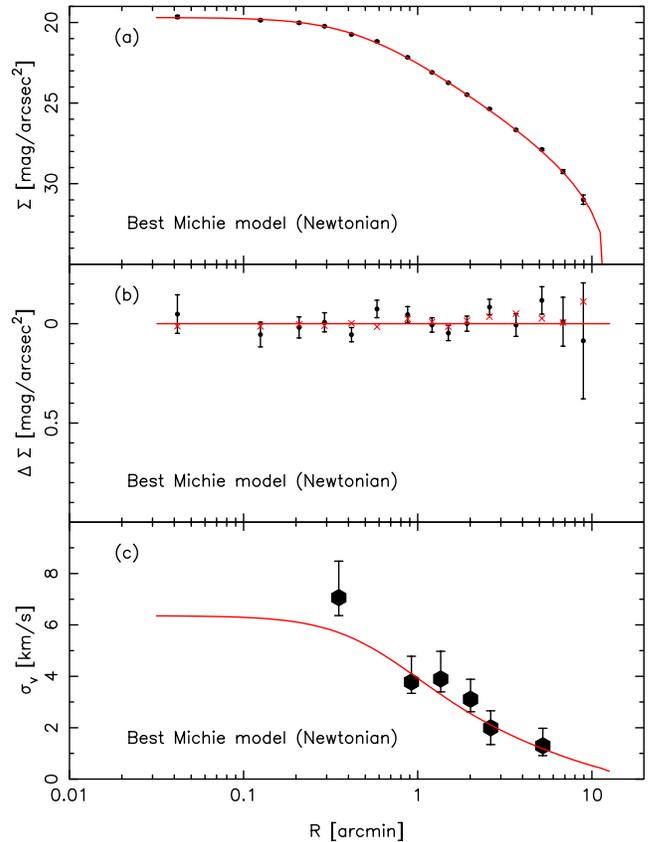}
\end{center}
\caption{As Figure~\ref{fig:best}, but showing the best Michie model derived in Paper~I, assuming Newtonian gravity (\#17 from that contribution).}
\label{fig:Michie}
\end{figure}

The surface brightness data, by their very nature, are binned averages. However, the kinematic data have much more discriminating power if they are left unbinned as individual line of sight velocity measurements. The likelihood of the model can then be calculated as the product of the likelihood of each data point (thus for these likelihood calculations we do not use the binned velocity dispersion estimates displayed on the bottom panels of Figures~\ref{fig:S11}-\ref{fig:Michie}, which are shown just to guide the eye). For the dynamical Michie models discussed in Paper~I, we integrated the distribution function to derive the line of sight velocity distribution as a function of radius. However, for the polytropic models analyzed here we did not attempt to derive the distribution function, as we judged this an unnecessary complication. The reason for this is that the observed line of sight velocity distribution closely resembles a Gaussian distribution at all radii (see Paper~I, Figure~14). We therefore assume that the radial and tangential velocity distributions are Gaussian, and expect this assumption to be a reasonable approximation to reality. Furthermore, it is unlikely that any small deviations from Gaussianity in the true line of sight velocity distributions will have any significant bearing on the results discussed below. It is straightforward then to integrate the polytropic model along the line of sight to obtain the predicted velocity dispersion at the projected radius of each kinematic datum. The full kinematic data set (samples A and B of Paper~I) are included in the analysis, since this gives the most favorable case for MOND.

The most likely model found in this manner has $c_0=7.7\kms$, $n=17.3$, $r_a=11.0\pc$ and $\rho_0=52.7 \msun \pc^{-3}$. By starting the MCMC algorithm from several different initial parameter combinations, we verified that this corresponds to a global likelihood maximum. However, this fit did not take into account whether the resulting best model is physically plausible. We checked that the models obey the Global Density Slope-Anisotropy Inequality \citep{Ciotti:2010p15374}, and checked also for stability by calculating $\xi_{half}$, a variant of the Fridman-Polyachenko-Shukhman parameter \citep{Fridman:1984p17072}, defined as twice the ratio of radial to tangential kinetic energy within the half-mass radius (so that $\xi_{half}=1$ for isotropic systems). \citet{Nipoti:2011p18855} have proposed $\xi_{half}$ as a stability indicator for spherical stellar systems in MOND, as the maximum value for stability $\xi_{half,s}$ is only weakly dependent on the density distribution and on the internal acceleration of the system: in particular, for MOND models of NGC~2419 in Paper I we found $\xi_{half,s} \sim 1.4-1.5$. The best model above has $\xi_{half}=1.53$, slightly beyond the stable region. Implementing a prior that forces $\xi_{half}<1.5$ in the fitting procedure, yields a model with parameters close to the previous best-fit, having $c_0=7.9\kms$, $n=17.6$, $r_a=11.5\pc$ and $\rho_0=58.0 \msun \pc^{-3}$, which has a total mass of $7.7\times 10^5 \msun$. This best (stable) polytropic model is displayed in Figure~\ref{fig:best}. The residuals with respect to the surface brightness profile are very much smaller than for the S11 model, and a first visual impression is that this is a reasonable model of the cluster. Nevertheless, the likelihood of this fit is a factor of $\Lambda=1/5035$ lower than the best Michie model fit in Paper~I (and shown in Figure~\ref{fig:Michie}): the surface brightness and kinematic data contribute a factor of 81 and 62 to this likelihood ratio, respectively\footnote{The highest likelihood fit ignoring the stability criterion has $\Lambda=1/4538$.}. In this comparison we have included an 8\% component of binaries to the Newtonian model, as fit in Paper~I; this has the effect of producing low-level ``wings'' to the velocity distribution. For the MOND polytropic model, the highest likelihood occurs with 0\% binaries.

Both the polytrope and Michie models have the same number of parameters: three structural parameters, plus an anisotropy radius, plus a fitted mass to light ratio and a velocity zero-point. Given that the fits to both models have an identical number of degrees of freedom, the quantity $-2\ln{\Lambda}$ will be asymptotically distributed as $\chi^2(1)$ \citep{James:2006p18770}. This allows us to exclude the best MOND polytrope at the 99.996\% confidence level.

While we judge the likelihood approach above to provide the strongest and most reliable method for model comparison, some readers may prefer the traditional frequentist $\chi^2$ hypothesis test. In the present context, the $\chi^2$ method has the disadvantage that it requires the kinematic data to be grouped into radial bins to derive the velocity dispersion. Since the individual velocity measurements come from different radial positions and have different uncertainties, much information is lost in calculating the sub-sample moments. Furthermore, the (weighted) velocity dispersion estimates listed in Paper~I have strongly asymmetric non-Gaussian uncertainties. The following $\chi^2$ values are therefore presented with these caveats. For the best MOND polytrope, we calculate from the 15 surface brightness data points $\chi^2_{SB}=22.55$, whereas the best Newtonian Michie model has $\chi^2_{SB}=13.75$. From the 6 velocity dispersion measurements (shown in the bottom panels of Figures~\ref{fig:S11}-\ref{fig:Michie}), we find for the MOND polytrope $\chi^2_{kin}=7.88$, and for the Newtonian Michie model $\chi^2_{kin}=6.76$. There are a total of 21 data points. The number of degrees of freedom for both models\footnote{Here we use the simpler Newtonian model without binaries.} is 15. The probability of an experiment with 15 degrees of freedom yielding $\chi^2=30.43$ (the value for the MOND polytrope fit) or greater by chance is 1\%. Hence the $\chi^2$ test (which for the reasons stated above is a weak statistical test for our kinematic data) allows us to reject the best polytropic model in MOND at the 99\% confidence level. In contrast, a value of $\chi^2=20.51$ (the case for the best Newtonian Michie model) occurs by chance with 15\% probability, which is perfectly acceptable. Inspection of Figure~\ref{fig:best}a shows that the last point in the surface brightness profile contributes significantly to the model discrepancy. Although there is no reason to doubt the validity of this datum, if we ignore it, the MOND polytropic model can still be rejected with 97\% confidence.

We note finally that it would have been possible to perform all the analysis described above by first de-projecting the surface brightness distribution to estimate the three-dimensional density $\rho(r)$, as was suggested to us by the anonymous referee; the model comparison would then have rested purely on the kinematic measurements. However, the significant uncertainties in the surface brightness profile at large radius mean that there is no single unequivocal solution to $\rho(r)$, and any uncertainty estimates of this function would involve problems of correlated noise. Hence our choice to project the models into the space of observables should be viewed as a statistically simpler option.

\section{Conclusions}

In the analysis presented above we have derived the highest likelihood polytropic model of NGC~2419, fitting to the observed kinematics and structure of this globular cluster assuming MOND and allowing also for the possibility of anisotropy in the stellar orbits. The model is compared to a previously-fitted model in Newtonian gravity. NGC~2419 is probably the best target for this analysis, since its extreme Galactocentric distance ($87.5\kpc$, \citealt{DiCriscienzo:2011p15907}) means that it is relatively unaffected by the external field due to the Milky Way (as confirmed by the N-body experiments presented in Paper I), while its large mass and luminosity allow us to resolve the radial velocity dispersion profile with a useful sample of stars. However, we have found that this best-fit polytropic model in MOND is a factor of $\sim 5000$ less likely than the best Michie model fit in Newtonian gravity, and can be rejected with high confidence.

S11 suggests that a MOND model deviating slightly from the polytropic relation might improve the fit. However, we would like to point out that in MOND as in Newtonian gravity, given the density profile of a spherical model, its velocity dispersion profile is fully determined by the anisotropy profile (in other words two different distribution functions generating the same density profile give different velocity dispersion profiles only if the corresponding anisotropy profiles are different). Thus given an observed surface brightness profile, for a choice of ${\rm M/L}$ and $\beta(r)$, there is only one possible velocity dispersion profile for each theory of gravity. In this sense, the experiment undertaken in Paper~I, where we searched through a vast number of possible stable anisotropies at different ${\rm M/L}$, is the definitive test to evaluate the relative likelihood of models. The fact that the polytropic models in MOND fare badly in comparison to the Newtonian Michie model was anticipated by that experiment. Furthermore, it must be noted that the best MOND model found with the MCMC analysis in Paper~I, which was shown to perform poorly with respect to the best Newtonian Michie model (being less likely by a factor of $\sim $ 350), turns out to be more likely than the S11 model (by a factor of $\sim 10^{238}$), but also than the best MOND polytrope studied here (by a factor of $\sim 14$). It follows that Sanders' (2011) suggestion to use MOND polytropes, though interesting, did not lead to finding MOND models of NGC~2419 better than those already considered in Paper~I. These results reinforce the conclusions presented in Paper~I: unless NGC~2419 is significantly non-spherical (for which there is no evidence from the projected measures we have access to), or has a radially-dependent mass to light ratio (which \citealt{Dalessandro:2008p15030} argue against, based on the distribution of Blue Straggler Stars), or is out of equilibrium, it provides a very strong challenge to MOND theory. 

The alternative to this deduction, suggested by S11, is that it is unwise to rely on formal statistical tests given that our data and physical analysis may be ``plagued by systematic effects''. While we have made the upmost effort to try to understand and take into account the noise properties of the instruments employed in this study (see Paper~I), it is of course conceivable that random and systematic uncertainties bias our results. However, no such problems have yet been brought to light, and in the absence of such evidence (which we are aware is not evidence of absence!), we are unwilling to give up the objective tool that is statistical inference.

\acknowledgments

R.I. gratefully acknowledges support from the Agence Nationale de la Recherche though the grant POMMME (ANR 09-BLAN-0228). We acknowledge the CINECA Award N. HP10C2TBYB, 2011 for the availability of high performance computing resources and support. C.N. is supported by the MIUR grant PRIN2008. M.B. acknowledges support by INAF through the PRIN-INAF 2009 grant CRA 1.06.12.10 (PI: R. Gratton) and by ASI through contracts COFIS ASI-INAF I/016/07/0 and ASI-INAF I/009/10/0. We would like to thank Robert Sanders and Beno\^it Famaey for helpful discussions concerning these models, and the anonymous referee for their comments.

\end{document}